\documentclass[prd,aps,reprint,noeprint,superscriptaddress,amsmath, amssymb]{revtex4-1}
\usepackage{graphicx}
\usepackage{subfigure}
\usepackage{lineno}
\usepackage{epsfig}
\usepackage{multirow}
\usepackage{overpic}
\usepackage{color}
\usepackage{bm} 
\usepackage{xspace} 
\usepackage{dcolumn}
\usepackage{booktabs,tabularx}
\usepackage{array}
\usepackage{textcomp}
\usepackage[colorlinks,linkcolor=blue,urlcolor=blue,citecolor=blue]{hyperref}
\usepackage{tikz}

\usepackage[normalem]{ulem}

\usepackage{cleveref}

\newcommand{\chisq}{\chi^{2}}

\begin{document}
\normalsize
\parskip=5pt plus 1pt minus 1pt

\title{\boldmath Observation of Transverse Polarization and Determination of Psionic Form Factors of the $\Lambda$ Hyperon at $\sqrt{s}= 3.773$~GeV
}
\author{BESIII Collaboration}
\thanks{Full author list given at the end of the Letter}
\begin{abstract}
Using a data sample of $e^{+}e^{-}$ collision data corresponding to an integrated luminosity of 20.3~fb$^{-1}$ collected by the BESIII detector at the BEPCII collider, we present an observation of transverse polarization and a complete determination of the psionic form factors of the $\Lambda$ hyperon in $e^{+}e^{-}\to{\Lambda}\bar{\Lambda}$ decay with the entangled ${\Lambda}-\bar{\Lambda}$ pair at $\sqrt{s}=3.773$~GeV.
The relative phase between the psionic form factors is determined to be $\Delta\Phi=(1.53\pm0.36\pm0.03)$~rad with a significance of 5.5~$\sigma$ taking into account systematic uncertainty. This result indicates a non-zero phase between the transition amplitudes of the ${\Lambda}\bar{\Lambda}$ helicity states.
Additionally, we measure the angular distribution parameter and the modulus of the ratio between the psionic form factors to be $\eta=0.86\pm0.05\pm0.03$ and $R(s)=|G_{E}(s)/G_{M}(s)|=0.47\pm0.08\pm0.05$, where the first uncertainty is statistical and the second systematic.
\end{abstract}

\maketitle

Understanding the structure of baryons is a major goal of contemporary particle physics~\cite{Brodsky:1974vy,Geng:2008mf,Green:2014xba,Wang:2022zyc}.
In the context of Quantum Chromodynamics (QCD), electromagnetic form factors (EMFFs) serve as important observables that connect measurable quantities to theoretical predictions.
In the 1960s, Cabibbo and Gatto~\cite{Cabibbo:1961sz} first proposed that timelike EMFFs could be studied at $e^+e^-$ experiments by measuring the production cross sections of baryon-antibaryon pairs. A large amount of research has been carried out regarding nucleon and strange hyperon EMFFs in the timelike momentum transfer regions ($s>0$)~\cite{ParticleDataGroup:2024cfk}, where $s$ is the square of the center-of-mass (c.m.) energy.
Among them, the proton, being a stable particle, can serve as a suitable target for studying its spacelike EMFFs through scattering experiments. A more recent study also revealed that for large $s$, the $d$-quark contributions to the proton EMFFs are reduced relative to the $u$-quark contributions~\cite{Cates:2011pz}.
In contrast to the proton, unstable hyperons with finite lifetimes are not suitable for such scattering experiments. Instead, the interaction in $e^+e^-$ collisions allows access to timelike EMFFs of hyperons due to virtual photon production in the $e^+e^-$ interaction which facilitates the quantitative assessment of the electromagnetic structure. Experimentally accessible timelike EMFFs are connected with the spacelike quantities, such as charge and magnetization densities, through the dispersion relation~\cite{Belushkin:2006qa}.
The pair production of spin-$1/2$ baryons can then be parametrized by the electric form factor $G_{E}(s)$ and the magnetic form factor $G_{M}(s)$~\cite{Wang:2022zyc,Qian:2021neg}, which are analytic functions of the momentum transfer squared.
In the timelike region, the EMFFs are complex and have a relative phase $\Delta\Phi=\Delta\Phi_{E}-\Delta\Phi_{M}$, i.e. $G_{E}(s)/G_{M}(s)= Re^{i\Delta\Phi}$ with the modulus of the ratio of EMFFs $R =|G_{E}(s)/G_{M}(s)|$.
This relative phase $\Delta\Phi$ reflects interfering production amplitudes and has a polarizing effect on the final state even if the initial state is unpolarized~\cite{Dubnickova:1992ii}.
This provides a handle to study the asymptotic properties of the timelike EMFFs related to the intrinsic structure of hyperons at large $s$, where the spacelike and timelike EMFFs should converge to the same value. 
For protons, the onset of this scale can be studied by measuring spacelike and timelike EMFFs, but it is hard to access to polarisation in these experiments.
For ground-state hyperons, on the other hand, the weak parity-violating decays provide straightforward access to their polarization and allow the measurement of the relative phase in the timelike region.

Experimentally, the first determination of the effective form factor of the $\Lambda$ hyperon was reported by the BABAR experiment using the initial state radiation (ISR) method~\cite{Aubert:2007uf}. Subsequently, Dobbs {\it et al} reported the measurements of the timelike EMFFs of several baryons~\cite{Dobbs:2014ifa,Dobbs:2017}.
Their conclusions regarding EMFFs and diquark correlations~\cite{Jaffe:2003sg} rely on the assumption that one-photon exchange dominates the production process and that contributions to the production through charmonium resonances are negligible. The spin formalism introduced in Ref.~\cite{Faldt:2017kgy} is also appropriate in the vicinity of vector charmonia. 
In this case, the form of the hadron current matrix element for the charmonia process is the same as for the virtual photon process. 
However, the form factors determined on-resonance receive contributions from charmonium decays, both electromagnetic and through the strong interaction, in addition to the purely electromagnetic one-photon exchange.
Recently, the BESIII collaboration performed a pioneering measurement of the relative phase, the modulus of the timelike EMFFs ratio and spin polarizations of the hyperons near threshold~\cite{BESIII:2019nep,BESIII:2020uqk,BESIII:2022kzc,BESIII:2023ynq}, around the resonances of vector charmonia~\cite{BESIII:2018cnd,BESIII:2020fqg,BESIII:2021ypr,BESIII:2022qax,BESIII:2022lsz,BESIII:2023lkg, BESIII:2023jhj,BESIII:2024nif,Liu:2023xhg,BESIII:2024dmr} and above the open charm threshold~\cite{BESIII:2019cuv,BESIII:2021ccp,BESIII:2021cvv,BESIII:2023rse,BESIII:2021cvv, BESIII:2024umc} by considering quantum entanglement of the hyperon and antihyperon.
This resulted in increased activity within the theoretical community, encompassing a variety of approaches including hyperon-antihyperon final state interactions~\cite{Haidenbauer:2020wyp,Qian:2022whn}, vector meson dominance~\cite{Yang:2019mzq,Li:2021lvs,Yan:2023yff,Chen:2023oqs}, the covariant spectator model~\cite{Ramalho:2019koj,Ramalho:2024wxp}, and dispersive calculations~\cite{Lin:2022baj,Mangoni:2021qmd}. 
In this Letter,  we report an observation of transverse polarization and determination of timelike psionic form factors of the $\Lambda$ hyperon using a multi-dimensional angular distribution analysis with a complete decomposition of the spin structure of the mixing process of $e^{+}e^{-}\to{\gamma^{*}/\psi}\to\Lambda\bar\Lambda$.
Note that the reaction in this work is a mixture of the virtual photon process and the potential $\psi(3770)$ resonance. In other words, the EM interaction is inherently involved. The measured form factor is a mixture of the EM and psionic form factors.

A data set corresponding to an integrated luminosity of 20.3~fb$^{-1}$ collected at $\sqrt{s}=3.773$~GeV~\cite{Ablikim:2014gna,BESIII:2024lbn} by the BESIII detector~\cite{Wang:2007tv} at the BEPCII collider~\cite{BESIII} is analyzed, which is about seven times larger than the one used in the previous study~\cite{BESIII:2021cvv}. To determine the detection efficiency and perform the unbinned maximum likelihood fit, $10^7$ Monte Carlo (MC) simulated events are generated using \textsc{kkmc}~\cite{kkmc, kkmc-01}, which includes the ISR effect.
The $e^+e^-\to\Lambda\bar{\Lambda}$ and ${\Lambda}(\bar{\Lambda})$ to $p\pi^{-}(\bar{p}\pi^{+})$ decays are simulated according to a phase space (PHSP) model using \textsc{evtgen}~\cite{evt2, evt2-01}. The response of the BESIII detector is modeled using a framework based on \textsc{geant}{\footnotesize 4}~\cite{geant4,geant4-01}.

\begin{figure}[!htbp]
\includegraphics[width=0.45\textwidth]{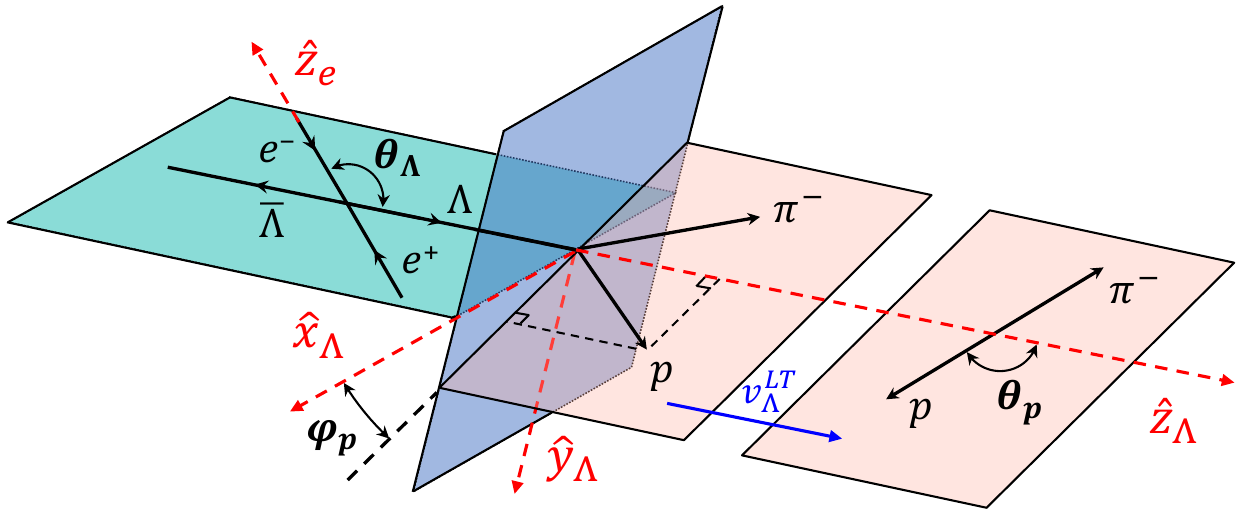}
{\color{blue}\put(-210, 35){$e^+e^-$}}
{\color{blue}\put(-213, 27){c.m.}}
{\color{blue}\put(-58, 63){$\Lambda$ rest frame}}
\caption{
Coordinate system utilized to describe the $e^{+}e^{-}\to{\Lambda}\bar{\Lambda}\to{p\bar{p}\pi^{+}\pi^{-}}$ process. The $\hat{z}_e$ axis is defined as the direction of $e^+$. The $\hat{z}_{\Lambda/\bar{\Lambda}}$ axis is determined as the direction of the $\Lambda/\bar{\Lambda}$ particle emission. The $\hat{y}_{\Lambda/\bar{\Lambda}}=\hat{z}_e\times\hat{z}_{\Lambda/\bar{\Lambda}}$, and $\hat{x}_{\Lambda/\bar{\Lambda}}=\hat{y}_{\Lambda/\bar{\Lambda}}\times\hat{z}_{\Lambda/\bar{\Lambda}}$. The $\theta_{\Lambda}$ is the angle between the $\Lambda$ hyperon and $e^+$ in $e^+e^-$ c.m. The angles $\theta_{p/\bar{p}}$ and $\phi_{p/\bar{p}}$ are the polar and azimuthal angles of the $p/\bar{p}$ momentum direction in the $\Lambda/\bar{\Lambda}$ rest frame, respectively. The $v_{\Lambda/\bar{\Lambda}}^{LT}$ is the Lorentz transformation with the velocity of $v_{\Lambda/\bar{\Lambda}}$.
}
\label{fig:helicity_frame}
\end{figure}

To describe the process of $e^{+}e^{-}\to{\Lambda}\bar{\Lambda}\to{p\bar{p}\pi^{+}\pi^{-}}$, it is essential to acquire information about each particle in the coordinate system of the parent particle. 
Here, a right-handed coordinate system is used according to that provided in Ref.~\cite{zhangzhe} to describe hyperon decays and the orientation of $p/\bar{p}$, see Fig.~\ref{fig:helicity_frame}.

In Ref.~\cite{zhangzhe}, the timelike EMFFs ratio $(R)$, the relative phase $\Delta\Phi$ and the angular distribution parameter $\eta$ are used to describe the process $e^{+}e^{-}\to{\gamma}^{*}/\psi\to{\Lambda}\bar{\Lambda}$. For $e^{+}e^{-}\to{\gamma^{*}/\psi}\to{\Lambda}\bar{\Lambda}\to{p\bar{p}\pi^{+}\pi^{-}}$, the joint decay angular distribution of this process is expressed in terms of the parameters of $\Delta\Phi$ and $\eta$ as
\begin{linenomath*}
\begin{equation}
\begin{aligned}
\label{w}
&\mathcal{W}(\bm\xi;\bm\Omega) = \mathcal{F}_{0}(\xi) + \eta\mathcal{F}_{5}(\xi)  \\
& \hspace{1em} + \alpha_{\Lambda}\alpha_{\bar{\Lambda}}[\mathcal{F}_{1}(\xi) + \sqrt{1-\eta^{2}}\cos(\Delta\Phi)\mathcal{F}_{2}(\xi) + \eta\mathcal{F}_{6}(\xi)] \\
& \hspace{1em} + \sqrt{1-\eta^{2}}\sin(\Delta\Phi)[\alpha_{\Lambda}\mathcal{F}_{3}(\xi) + \alpha_{\bar{\Lambda}}\mathcal{F}_{4}(\xi)],
\end{aligned}
\end{equation}
\end{linenomath*}
Here, $\boldsymbol\Omega~=~\left(\eta,~\Delta\Phi,~\alpha_{{\Lambda}},~\alpha_{\bar{\Lambda}}\right)
$ is the polarization parameters and the $\xi~=~\left(\theta_{\Lambda},~\theta_{p},~\theta_{\bar{p}},~\phi_{p},~\phi_{\bar{p}}\right)
$ is the measured angles defined in Fig.~\ref{fig:helicity_frame}. 
The angular functions $\mathcal{F}_{j}(\xi)$ $(j=0,1,...,6)$ are defined as
\begin{align}
 \label{Fn}
 {\cal{F}}_0 = &1,\nonumber\\
{\cal{F}}_1 = &\sin^2\!\theta_{\Lambda} \sin\!\theta_{p}\sin\!\theta_{\bar{p}}\cos\!\phi_{p}\cos\!\phi_{\bar{p}} - \cos^2\!\theta_{\Lambda} \cos\!\theta_{p}\cos\!\theta_{\bar{p}},\nonumber\\
{\cal{F}}_2 = &\sin\!\theta_{\Lambda}\cos\!\theta_{\Lambda} (\sin\!\theta_{p}\cos\!\theta_{\bar{p}}\cos\!\phi_{p} - \cos\!\theta_{p}\sin\!\theta_{\bar{p}}\cos\!\phi_{\bar{p}}),\nonumber\\
{\cal{F}}_3 = &-\sin\!\theta_{\Lambda}\cos\!\theta_{\Lambda} \sin\!\theta_{p}\sin\!\phi_{p},\\
{\cal{F}}_4 = &\sin\!\theta_{\Lambda}\cos\!\theta_{\Lambda} \sin\!\theta_{\bar{p}}\sin\!\phi_{\bar{p}},\nonumber\\
{\cal{F}}_5 = &\cos^2\!\theta_{\Lambda},\nonumber\\
{\cal{F}}_6 = &\sin^2\!\theta_{\Lambda} \sin\!\theta_{p}\sin\!\theta_{\bar{p}}\sin\!\phi_{p}\sin\!\phi_{\bar{p}} - \cos\!\theta_{p}\cos\!\theta_{\bar{p}}.\nonumber
\end{align}
Note that the difference for the $F_{i}(\xi)$ definition in Eq.(~\ref{Fn}) compared to the previous work~\cite{BESIII:2021cvv} is due to the different choice of coordinate system in Ref.~\cite{zhangzhe} and Ref.~\cite{Faldt:2017kgy}.
Furthermore, $\alpha_{\Lambda(\bar{\Lambda})}$ represents the decay parameters of $\Lambda(\bar{\Lambda})\to p\pi^{-}(\bar{p}\pi^{+})$ and $\eta$ denotes the scattering angle distribution parameter related to $R$ by
\begin{align}
    \label{eta}
    \eta=\frac{\tau-R^2}{\tau+R^2},
\end{align}
with $\tau=s/4m^{2}_{\Lambda}$. 
Since the production process is either strong or electromagnetic and thus parity conserving, if the initial state is unpolarized, non-zero transverse polarization can only occur in the direction normal to
the production plane, i.e., the $y$ direction. 
The hyperon transverse polarization is defined as 
\begin{align}
    \label{R}
    P_{y}=\frac{\sqrt{1-\eta^{2}}\sin\theta_{\Lambda}\cos\theta_{\Lambda}}{1+\eta\cos^{2}\theta_{\Lambda}}\sin(\Delta\Phi).
\end{align}

Charged tracks are reconstructed in the multi-layer drift chamber within its angular coverage, $|\cos\theta|<0.93$, where $\theta$ is the polar angle with respect to the $e^{+}$ beam direction in the laboratory system. The numbers of negatively and positively charged tracks of events are both larger than one. Charged tracks with momenta greater than 0.6~GeV/$c$~\cite{BESIII:2023euh} are identified as $p(\bar{p})$, while others are assigned as $\pi^{+}(\pi^{-})$.
 
For further analysis, fully reconstructed $e^{+}e^{-}\to{\Lambda}\bar{\Lambda}$ events with $\Lambda\to{p\pi^{-}}$ and $\bar{\Lambda}\to{\bar{p}\pi^{+}}$ are selected. To reconstruct $\Lambda(\bar\Lambda)$ candidates, a vertex fit and a secondary vertex fit~\cite{XUM} are applied to all combinations of one $p(\bar{p})$ track and one $\pi^{-}(\pi^{+})$ track. 
From all possible combinations, the one with the minimum value of $\sqrt{|M_{p\pi^{-}}-m_{\Lambda}|^{2} + |M_{\bar{p}\pi^{+}}-m_{\bar\Lambda}|^{2}}$ is selected. Here, $M_{p\pi^{-}(\bar{p}\pi^{+})}$ denotes the invariant mass of the $p\pi^{-}(\bar{p}\pi^{+})$ pair, and $m_{\Lambda(\bar\Lambda)}$ represents the nominal mass of $\Lambda(\bar\Lambda)$~\cite{ParticleDataGroup:2024cfk}.
To further suppress background contributions from non-$\Lambda(\bar\Lambda)$ events, the decay lengths of $\Lambda$ and $\bar\Lambda$ are both required to be greater than zero, where the negative decay lengths are due to detector resolution.

\begin{figure}[!htbp]
\includegraphics[width=0.45\textwidth]{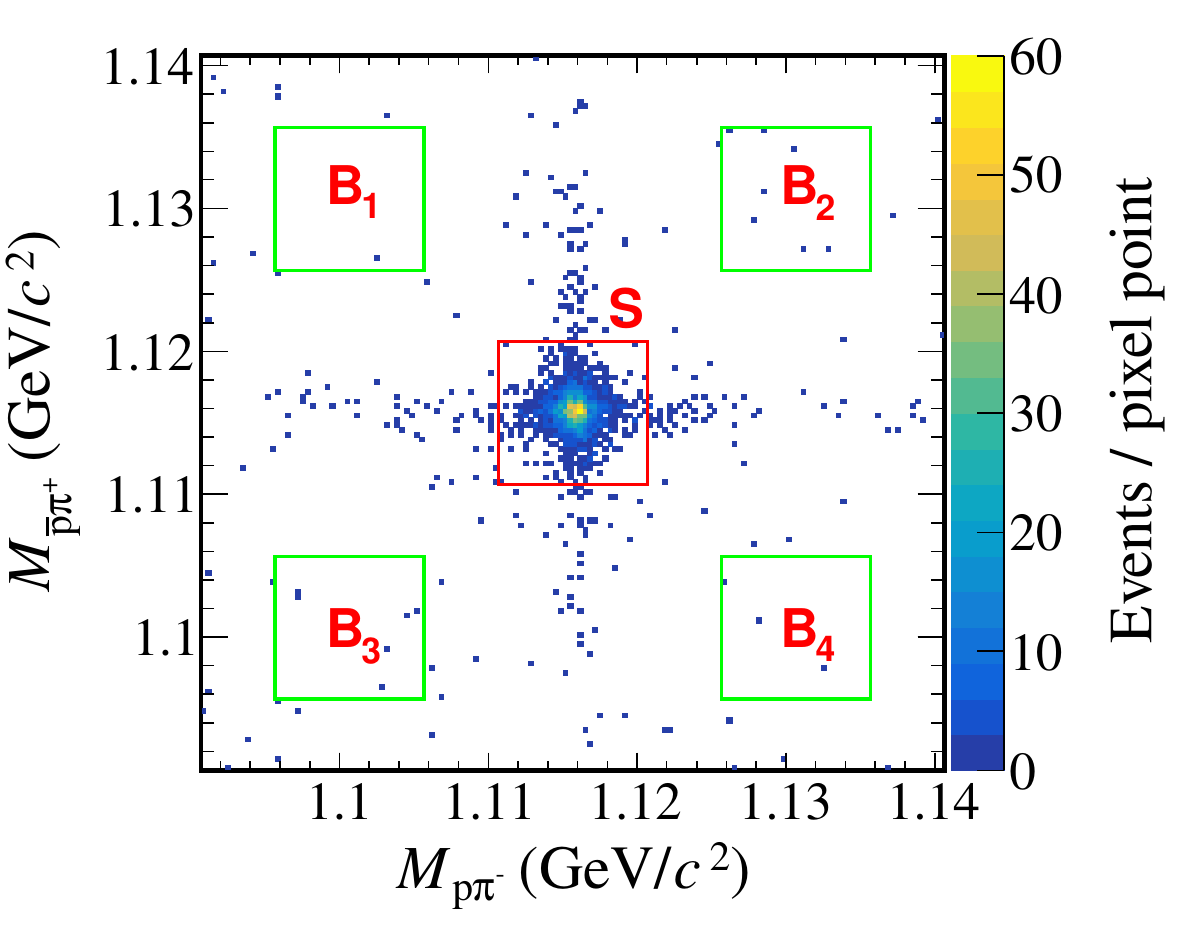}
\caption{Two-dimensional distribution of $M_{\bar{p}\pi^{+}}$ versus $M_{p\pi^{-}}$ for data, where the red box indicates the signal region, and the green boxes show the selected sideband regions.
}
\label{scatter_plot::llb}
\end{figure}

 After $\Lambda(\bar{\Lambda})$ reconstruction, a four-constraint (4C) kinematic fit is applied to all $\Lambda(\bar{\Lambda})$ hypotheses, enforcing energy-momentum conservation from the initial $e^{+}e^{-}$ to the final $\Lambda(\bar{\Lambda})$ state and combined with the requirement of $\chisq_{\rm{4C}}<100$. 
 Figure~\ref{scatter_plot::llb} shows the distribution of $M_{\bar{p}\pi^{+}}$ versus $M_{p\pi^{-}}$ after applying all above selection.
 The invariant mass of $p\pi^{-}(\bar{p}\pi^{+})$ is required to be within 5~MeV/$c^{2}$ of the $\Lambda(\bar\Lambda)$ mass taken from~\cite{ParticleDataGroup:2024cfk} ($|M_{p\pi^{-}(\bar{p}\pi^{+})}-m_{\Lambda(\bar\Lambda)}|<$ 5~MeV/$c^{2}$).
The signal region, denoted by $S$ in Fig.~\ref{scatter_plot::llb}, is determined and optimized using the figure of merit ${S}/{\sqrt{S + B}}$ derived from MC sample. Here, $S$ represents the number of signal MC events, and $B$ corresponds to the expected number of background events from the inclusive MC simulation of $e^{+}e^{-}\to$ hadron events. The background comes from non-$\Lambda(\bar\Lambda)$ events, such as $e^{+}e^{-}\to\pi^{+}\pi^{-}p\bar{p}$, that can be estimated using the corner method, {\it i.e.,} $\sum^{4}_{i=1}B_{i}/4$ for $M_{p\pi^{-}}$ and $M_{\bar{p}\pi^{+}}$ windows.
Here, the definition for regions ($B_{i}$) displayed in Fig.~\ref{scatter_plot::llb} is the same as the one used in Ref.~\cite{BESIII:2021cvv}. 
The number of background events estimated from the aforementioned corner method is $4\pm2$, which is negligible.
To estimate peaking background contributions such as $e^+e^-\to\gamma\psi(3686)\to\gamma\Lambda\bar{\Lambda}$, an inclusive MC sample of $\psi(3770)$ is employed. The number of background events of ISR $\psi(3686)$ is estimated to be $39\pm6$ events with a background level of approximately $1.8\%$ of the signal yield.
Note that the possible backgrounds for $\Sigma^{0}\to\gamma\Lambda$ and $K^{0}_{S}$ decays are highly suppressed based on the event selection criteria introduced above.
The number of observed events in data is determined to be  $2194\pm48$ with a signal MC efficiency of (37.00$\pm$0.03)\%.

\begin{figure}[!htbp]
\includegraphics[width=0.45\textwidth]{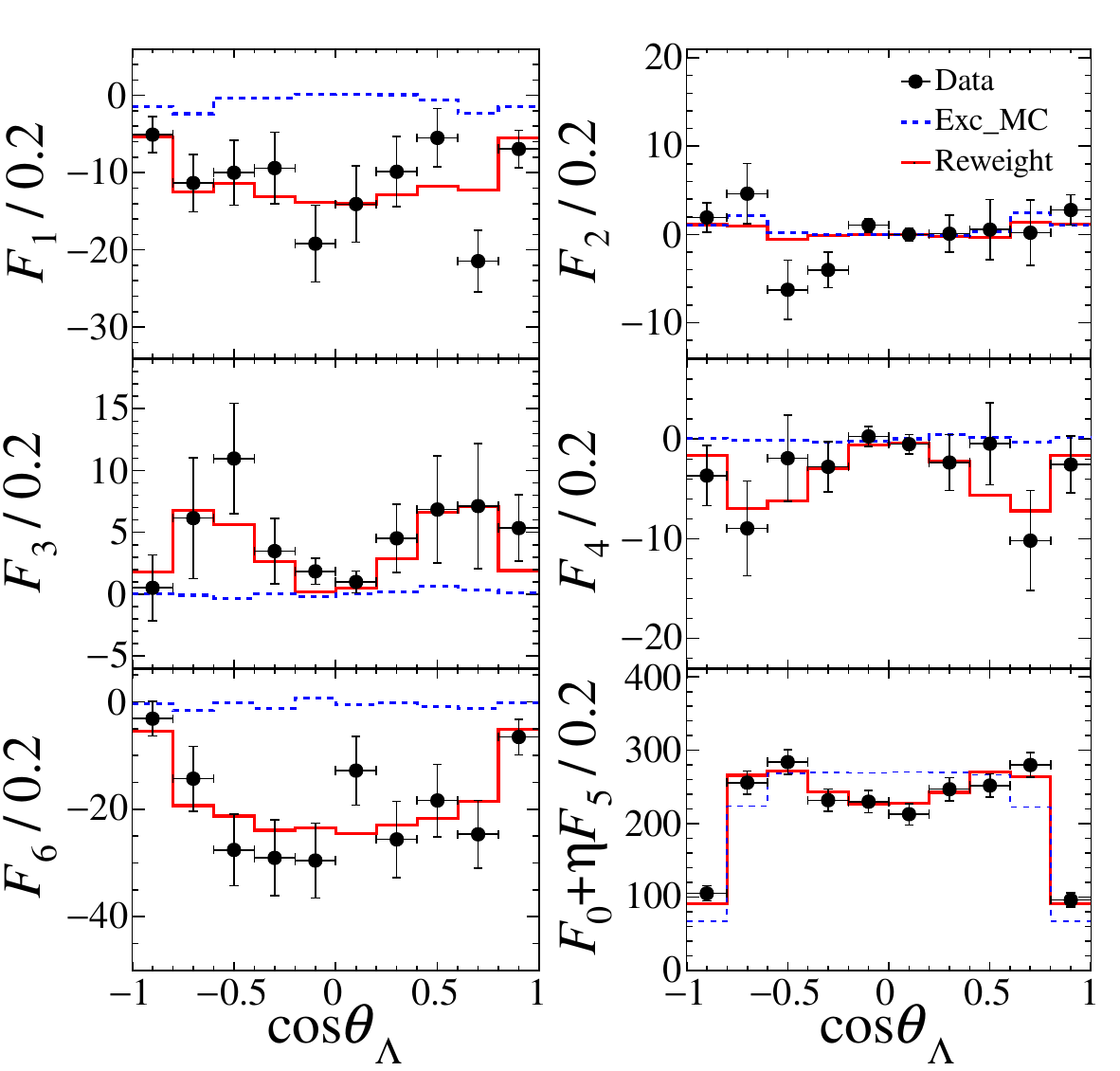}
\caption{The $F_{k} (k = 1, 2, 3, 4, 6)$ moment and $F_{0} +\eta F_{5}$ distribution (last panel) with respect to $\cos\theta_{\Lambda}$. The dots with error bars are the data, and the red line is the weighted PHSP MC corrected by the results of the global fit in Eq.~(\ref{Sss}). The blue dashed line is the distributions for unweighted simulated PHSP events. }
\label{scatter_plot::llb:projections}
\end{figure}

To determine the set of $\Lambda$ spin polarization parameters ($\Delta\Phi, \eta$), an unbinned maximum likelihood fit is performed. In the fit,  $\alpha_{\Lambda/\bar\Lambda}$ is fixed to $\pm0.7542$ by referring to~\cite{BESIII:2022qax} assuming charge-parity conservation.
The likelihood function $\mathcal{L}$ is constructed from the probability density function (PDF), ${\cal{P}}({\boldsymbol{\xi}}_i)$, for the event $i$ characterized by the measured angles $\boldsymbol{\xi}_i$ as
\begin{align}
    \label{L}
    \mathcal{L} =\prod_{i=1}^{N}\mathcal{P}(\boldsymbol\xi_{i},\mathbf{\Omega})=\prod_{i=1}^{N}\mathcal{C}\mathcal{W}(\boldsymbol\xi_{i},\mathbf{\Omega})\epsilon (\boldsymbol\xi_{i}),
\end{align}
where $N$ is the number of events in the signal region. The joint angular distribution ${\cal{W}}({\boldsymbol{\xi}}_i, {\boldsymbol{\Omega}})$ is given in Eq.~\eqref{w}, and $\epsilon(\boldsymbol{\xi}_i)$ is the detection efficiency.
The normalization factor $\mathcal{C}^{-1}=\frac{1}{N_\mathrm{MC}}\sum_{j=1}^{N_\mathrm{MC}} {\cal{W}}({\boldsymbol{\xi}}^{j}, {\boldsymbol{\Omega}})$ is calculated as a sum of the corresponding amplitudes $\cal{W}$ from the accepted PHSP MC events $N_\mathrm{MC}$, applying the same event selection criteria as to the data.
The minimization of the objective function defined as
\begin{align}
  \label{Sss}
\mathit{S} = -\mathrm{ln}\mathcal{L}_{\rm S} + \mathrm{ln}\mathcal{L}_{\rm B},
\end{align}
is conducted using the MINUIT package from the ROOT library~\cite{James:1975dr}.
In Eq.~\eqref{Sss}, $\mathcal{L}_{\rm S}$ and $\mathcal{L}_{\rm B}$ represent the likelihood function for events chosen in the signal region and sideband regions.
Figure~\ref{scatter_plot::llb:projections} shows the distributions of the five moments
$F_j$ ($j=1,2,3,4,6$). They are the projections of $\mathcal{F}_j$ ($j=1,2,3,4,6$) in Eq. (2) onto $\cos\theta$, defined as ${F}_{j} = \sum_i^{N}(\mathcal{F}_{j}(\xi_{i}))$. 
And the $\Lambda$ angular distribution (${{F}}_{0} +\eta{{F}}_{5}$) is analyzed with respect to $\cos\theta_{\Lambda}$ in 10 intervals.

\begin{figure}[!htbp]
 \includegraphics[width=0.45\textwidth]{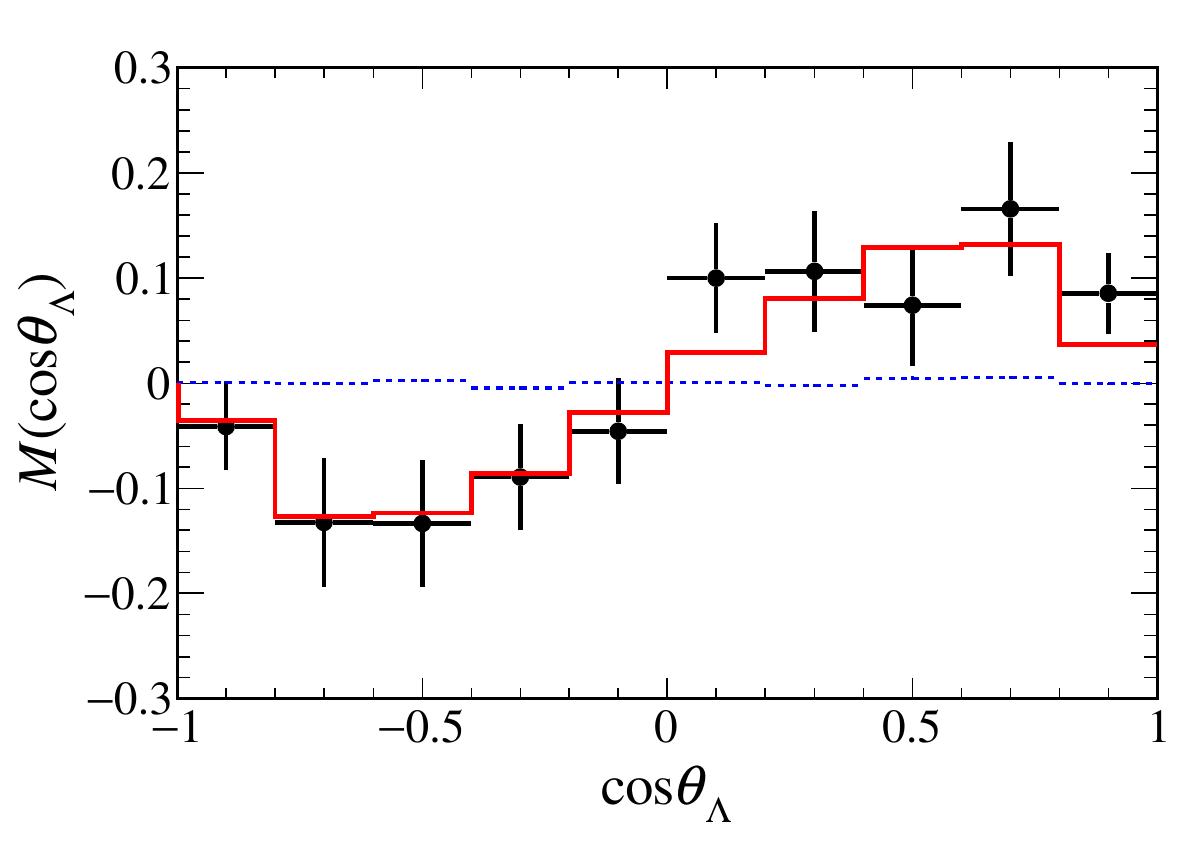}
\caption{The moments $M(\cos\theta_{\Lambda})$ as a function of $\cos\theta_{\Lambda}$.
The dots with error bars are data, and the red line is the weighted PHSP MC corrected by the results of the global fit in Eq.~(\ref{Sss}). The blue dashed line is the distributions from simulated PHSP events.
}
\label{scatter_plot::llb:polarization}
\end{figure}

The moment 
\begin{equation}\label{moment}
 M(\cos\theta_{\Lambda}) = -\frac{m}{N}\sum_i^{N_k}(\sin\!\theta_{p}^{i}\sin\!\phi_{p}^{i} + \sin\!\theta_{\bar{p}}^{i}\sin\!\phi_{\bar{p}}^{i}),
\end{equation}
 is related to the transverse polarization and calculated for $m = 10$ intervals in $\cos\theta_{\Lambda}$. Here, $N_k$ denotes the number of events in the $k$th $\cos\theta_{\Lambda}$ interval. And $N=\sum_k^{m}N_k$ is the number of observed events in the data set.
The expected angular dependence of the moment for the
acceptance-corrected data reads
\begin{align}
\label{moments}
M(\cos\theta_{\Lambda})&=(\alpha_{\Lambda}-\alpha_{\bar\Lambda}) \frac{\sqrt{1-\eta^{2}}}{3+\eta}\sin\Delta\Phi\cos\theta_{\Lambda}\sin\theta_{\Lambda} \\
&=(\alpha_{\bar\Lambda}-\alpha_{\Lambda})\frac{{1+\eta\cos^{2}\theta_{\Lambda}}}{3+\eta}P_{y}.
\end{align}
Figure~\ref{scatter_plot::llb:polarization} shows the distribution of the $M(\cos\theta_{\Lambda})$ defined in Eq.~\eqref{moment}, reflecting the fit result. 
It is consistent with the behavior described by Eq.~\eqref{R} when compared to the data.
The significance of the transverse polarization signal, considering systematic uncertainties, is determined to be 5.5$\sigma$ by comparing the likelihoods with and without transverse polarization. 

Systematic uncertainties in the measurement of $\Lambda$ hyperon polarization arise from various sources, including background contributions, $\Lambda$ reconstruction, kinematic fit, beam transverse polarization, decay parameters of $\Lambda\to p\pi$, and the fit method. The background contributions include the sideband region which describes the non-$\Lambda(\bar\Lambda)$ background and ISR $\psi(3686)$ background. The uncertainty due to background candidates is estimated by comparing the fits with and without the background contributions, where the contribution from the ISR $\psi(3686)$ background is the dominant one.
To estimate the uncertainty related to the $\Lambda$ reconstruction including the tracking, the requirement on the mass window and decay length of $\Lambda$, it is studied from a control sample of $\psi(3686)\to\Lambda\bar{\Lambda}$ events.
The uncertainty arising from this source is evaluated using the same method as in Ref.~\cite{BESIII:2022lsz}. The discrepancy between the nominal and average values obtained from variations is taken as the systematic uncertainty.
The uncertainty associated with the kinematic fit is assigned as the results with and without track helix parameter corrections~\cite{ref5}.
The systematic uncertainty originating from the transverse beam polarization is estimated by changing the joint decay angular distribution $\mathcal{W}$ according to Ref.~\cite{shuliu}. The difference of results between the nominal and released parameters of transverse beam polarization is taken as the systematic uncertainty.
The uncertainty caused by the fixed decay parameters of $\alpha_{\Lambda/\bar\Lambda}$ is estimated by varying mean values obtained from averaging results in~\cite{BESIII:2022qax} within $\pm1\sigma$. The change of the result is negligible and thereby the related uncertainty is neglected.
The reliability of the fit results is validated by performing an input and output check based on 300 pseudoexperiments using the helicity amplitude formula from Ref.~\cite{BESIII:2022qax}.
The mean value of polarization parameters measured in the analysis ($\eta$=0.86, $\Delta\Phi$=1.53) are used as input in the formula, and the number of events in each generated MC sample is ten times of the data sample.
The difference between the input and output results is taken as the systematic uncertainty.
Assuming all sources to be independent, the total systematic uncertainty is calculated as the square root of their quadratic sum. All systematic uncertainties are listed in Table~\ref{tab:Xi:Rec:eff:polarization}.

\begin{figure}[!htbp]
  \includegraphics[width=0.48\textwidth]{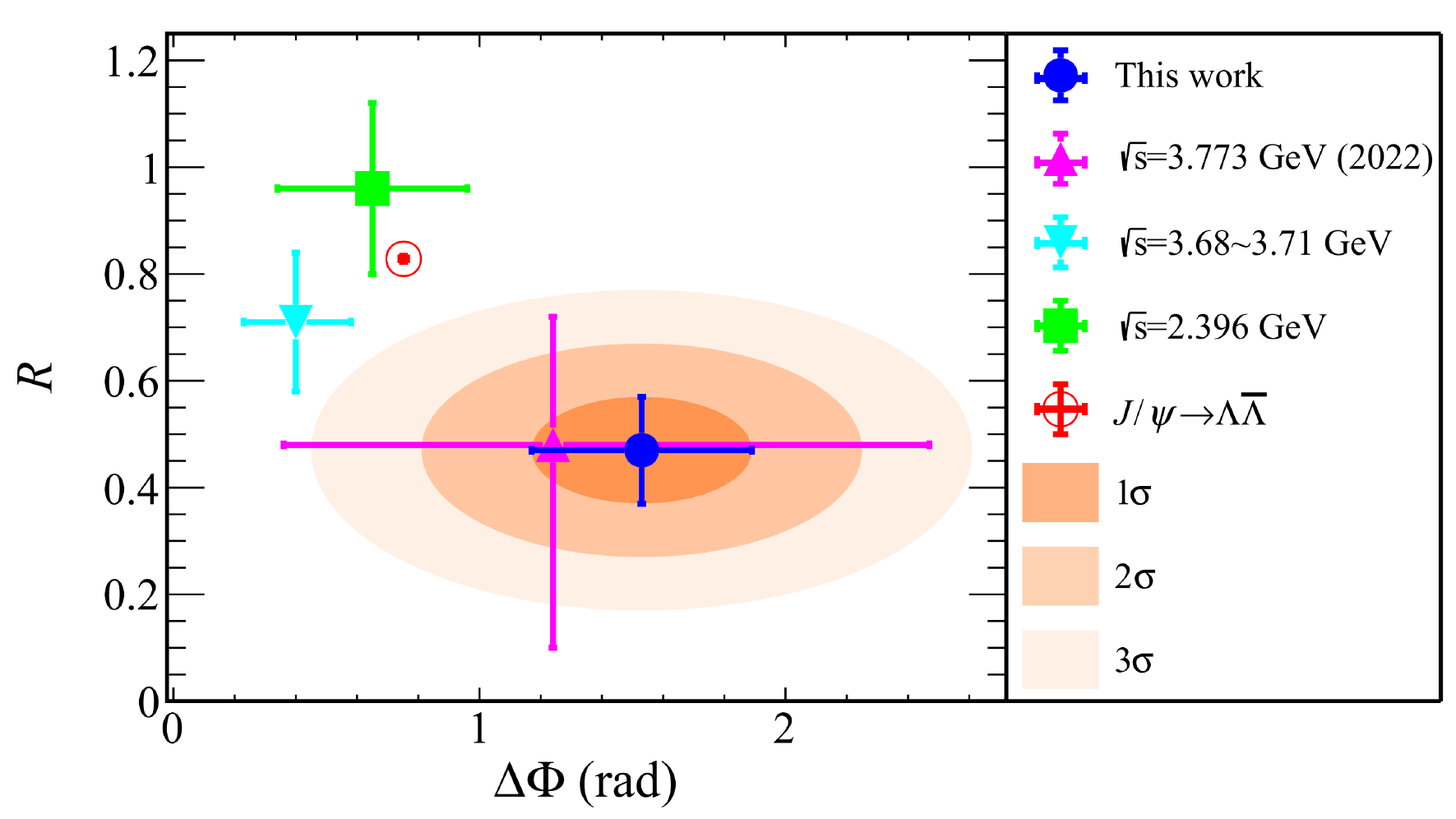}
\caption{
Two dimensional distribution of $\Delta\Phi$ and $R$ between this work and previous BESIII measurements at different c.m. energies~\cite{BESIII:2019nep,BESIII:2021cvv,BESIII:2022qax,BESIII:2023euh}. The uncertainty combines from both statistical and systematic uncertainties. 
The inner, intermediate and outer contours in orange represent 68.2\%, 95.4\%, and 99.7\% confidence level, respectively. }
\label{scatter_plot::llb:RPhi}
\end{figure}

\begin{table}[!hbpt]
\begin{center}
\caption{The absolute systematic uncertainties in the measurement of the $\Lambda$ hyperon polarization parameters.}
\begin{tabular}{lccc} \hline\hline
Source                                                   &$\eta$ &$\Delta\Phi$ (rad) &$R$ \\ \hline
				Backgrounds           		   	&{0.02}  	&{0.01}	&{0.04}\\
                 {$\Lambda$ reconstruction} 		&{0.02}	 	&{0.01}	&{0.04}\\
                 Kinematic fit      			   		&{0.00}	 	&{0.01}	&{0.00}\\
                 Beam transverse polarization           &{0.00}	 	&{0.02}	&{0.00}	 \\
                 Fit method				   			&{0.00}  	&{0.02}	&{0.00}\\
\hline
				 Total  					   			&{0.03}  	&{0.03} &{0.05}\\ \hline\hline
\end{tabular}
\label{tab:Xi:Rec:eff:polarization}
\end{center}
\end{table}
In summary, we report the observation of transverse polarization and
complete determination of timelike psionic form factors of the $\Lambda$ hyperon in the $e^{+}e^{-}\to{\Lambda}\bar{\Lambda}$ process at \linebreak $\sqrt{s} = 3.773$~GeV, using a data sample corresponding to an integrated luminosity of 20.3~fb$^{-1}$ collected by the BESIII detector, offering a higher precision compared with the previous measurement~\cite{BESIII:2021cvv}.
The relative phase, the angular distribution parameter and the modulus of the timelike psionic form factors ratio are measured to be \linebreak $\Delta\Phi=(1.53\pm0.36\pm0.03)$~rad,
$\eta=0.86\pm0.05\pm0.03$ and
$R=0.47\pm0.08\pm0.05$, respectively, where the first uncertainty is statistical and the second systematic. 
For the first time, we observe a relative phase that differs significantly from zero (at 5.5 $\sigma$, taking systematic uncertainties into account) in the high momentum transfer region ($s>14$~GeV$^2$).
A comparison of $\Delta\Phi$ and $R$ between this work and previous measurements at different c.m. energies~\cite{BESIII:2019nep,BESIII:2021cvv,BESIII:2022qax,BESIII:2023euh}
are illustrated in Fig.~\ref{scatter_plot::llb:RPhi}. 
The measured polarization in this work is consistent with and more precise than the previous measurements at $\sqrt{s} = 3.773$~GeV~\cite{BESIII:2022qax}, and $\Delta\Phi$ is also roughly consistent with the results at other c.m. energies with an uncertainty of $(1-2)\sigma$.
However, noticeable differences exist in the $R$ value between this work and other c.m. energies. In particularly, this measurement compared to the one at $\sqrt{s} = 2.396$ GeV below the region of charmonium production, indicates potential variations in production mechanisms
at different c.m. energies.

Spin-1/2 hyperons produced in a hyperon-antihyperon pair can have either the same or opposite helicity. The nonvanishing relative phase $\Delta\Phi$ between the transition amplitudes of these helicity states implies the contributions not only from the $S$-wave but also $D$-wave amplitudes to ${\Lambda}\bar{\Lambda}$ production.
Since the measured $R$ values suggest that an energy dependence of this value, more data samples at various c.m. energies are needed for a detailed study of the phase dependence on the momentum transfer squared, $s$.
The clear and prominent signal enhances our understanding of the ${\Lambda}\bar{\Lambda}$ production mechanism within the $e^{+}e^{-}\to\gamma^{*}/\Psi\to{\Lambda}\bar{\Lambda}$ 
process, providing valuable insights into the structure of baryons.

{\it Acknowledgement}---
The BESIII Collaboration thanks the staff of BEPCII (https://cstr.cn/31109.02.BEPC) and the IHEP computing center for their strong support. This work is supported in part by
the Fundamental Research Funds for the Central Universities under Contracts No. lzujbky-2025-ytA05,  No. lzujbky-2025-it06,  No. lzujbky-2024-jdzx06;
Guangdong Provincial Key Laboratory of Advanced Particle Detection Technology No. 2024B1212010005;
Guangdong Provincial Key Laboratory of Gamma-Gamma Collider and Its Comprehensiv Applications No. 2024KSYS001;
National Key R\&D Program of China under Contracts Nos. 2023YFA1606000, 2023YFA1606704; National Natural Science Foundation of China (NSFC) under Contracts No. 12075107, No. 12247101, No. 11635010, No. 11935015, No. 11935016, No. 11935018, No. 12025502, No. 12035009, No. 12035013, No. 12061131003, No. 12192260, No. 12192261, No. 12192262, No. 12192263, No. 12192264, No. 12192265, No. 12221005, No. 12225509, No. 12235017, No. 12361141819; 
the Natural Science Foundation of Gansu Province No. 22JR5RA389, No.25JRRA799;
by the ‘111 Center’ under Grant No. B20063;
the Chinese Academy of Sciences (CAS) Large-Scale Scientific Facility Program; CAS under Contract No. YSBR-101; 100 Talents Program of CAS; The Institute of Nuclear and Particle Physics (INPAC) and Shanghai Key Laboratory for Particle Physics and Cosmology; German Research Foundation DFG under Contract No. FOR5327; Istituto Nazionale di Fisica Nucleare, Italy; Knut and Alice Wallenberg Foundation under Contracts Nos. 2021.0174, 2021.0299; Ministry of Development of Turkey under Contract No. DPT2006K-120470; National Research Foundation of Korea under Contract No. NRF-2022R1A2C1092335; National Science and Technology fund of Mongolia; National Science Research and Innovation Fund (NSRF) via the Program Management Unit for Human Resources \& Institutional Development, Research and Innovation of Thailand under Contract No. B50G670107; Polish National Science Centre under Contract No. 2024/53/B/ST2/00975; Swedish Research Council under Contract No. 2019.04595; U. S. Department of Energy under Contract No. DE-FG02-05ER41374.

\begin{widetext}
\begin{center}
\small
M.~Ablikim$^{1}$, M.~N.~Achasov$^{4,c}$, P.~Adlarson$^{77}$, X.~C.~Ai$^{82}$, R.~Aliberti$^{36}$, A.~Amoroso$^{76A,76C}$, Q.~An$^{73,59,a}$, Y.~Bai$^{58}$, O.~Bakina$^{37}$, Y.~Ban$^{47,h}$
% ...
\\
\vspace{0.2cm}
(BESIII Collaboration)\\
\vspace{0.2cm} {\it
$^{1}$ Institute of High Energy Physics, Beijing 100049, People's Republic of China\\
% ...
$^{4}$ Budker Institute of Nuclear Physics SB RAS (BINP), Novosibirsk 630090, Russia\\
% ...
$^{36}$ Johannes Gutenberg University of Mainz, Johann-Joachim-Becher-Weg 45, D-55099 Mainz, Germany\\
$^{37}$ Joint Institute for Nuclear Research, 141980 Dubna, Moscow region, Russia\\
% ...
$^{47}$ Peking University, Beijing 100871, People's Republic of China\\
% ...
$^{58}$ Southeast University, Nanjing 211100, People's Republic of China\\
$^{59}$ State Key Laboratory of Particle Detection and Electronics, Beijing 100049, Hefei 230026, People's Republic of China\\
% ...
$^{73}$ University of Science and Technology of China, Hefei 230026, People's Republic of China\\
% ...
$^{76}$ University of Turin and INFN, (A)University of Turin, I-10125, Turin, Italy; (B)University of Eastern Piedmont, I-15121, Alessandria, Italy; (C)INFN, I-10125, Turin, Italy\\
$^{77}$ Uppsala University, Box 516, SE-75120 Uppsala, Sweden\\
% ...
$^{82}$ Zhengzhou University, Zhengzhou 450001, People's Republic of China\\

\vspace{0.2cm}
$^{a}$ Deceased\\
% ...
$^{c}$ Also at the Novosibirsk State University, Novosibirsk, 630090, Russia\\
% ...
$^{h}$ Also at State Key Laboratory of Nuclear Physics and Technology, Peking University, Beijing 100871, People's Republic of China\\
% ...
}
\end{center}
\end{widetext}

\end{document}